# STUDY PAPER ON TEST CASE GENERATION FOR GUI BASED TESTING

Isabella[1] and Emi Retna[2]

[1]PG Research Scholar, Software Engineering Group, School of Computer Science and Technology, Karunya University, Coimbatore, India
`aisabella20@gmail.com`
[2]Head – Computer Technology Centre, Karunya University, Coimbatore, India
`emiretna@gmail.com`

## ABSTRACT

*With the advent of WWW and outburst in technology and software development, testing the software became a major concern. Due to the importance of the testing phase in a software development lifecycle, testing has been divided into graphical user interface (GUI) based testing, logical testing, integration testing, etc. GUI Testing has become very important as it provides more sophisticated way to interact with the software. The complexity of testing GUI increased over time. The testing needs to be performed in a way that it provides effectiveness, efficiency, increased fault detection rate and good path coverage. To cover all use cases and to provide testing for all possible (success/failure) scenarios the length of the test sequence is considered important. Intent of this paper is to study some techniques used for test case generation and process for various GUI based software applications.*

## 1. INTRODUCTION

Graphical User Interface (GUI) is a program interface that takes advantage of the computer's graphics capabilities to make the program easier to use. Graphical User Interface (GUI) provides user an immense way to interact with the software [1]. The most eminent and essential parts of the software that is being used today are Graphical User Interfaces (GUIs) [8], [9]. Even though GUIs provides user an easy way to use the software, they make the development process of the software tangled [2].

Graphical user interface (GUI) testing is the process of testing software's graphical user interface to safeguard it meets its written specifications and to detect if application is working functionally correct. GUI testing involves performing some tasks and comparing the result with the expected output. This is performed using test cases. GUI Testing can be performed either manually by humans or automatically by automated methods.

Manual testing is done by humans such as testers or developers itself in some cases and it is often error prone and there are chances of most of the test scenarios left out. It is very time consuming also. Automated GUI Testing includes automating testing tasks that have been done manually before, using automated techniques and tools. Automated GUI testing is more, efficient, precise, reliable and cost effective.

A test case normally consists of an input, output, expected result and the actual result. More than one test case is required to test the full functionality of the GUI application. A collection of test cases are called test suite. A test suite contains detailed guidelines or objectives for each collection of test cases.

Model Based Testing (MBT) is a quick and organized method which automates the testing process through automated test suite generation and execution techniques and tools [11]. Model based testing uses the directed graph model of the GUI called event-interaction graph (EIG) [4] and event semantic interaction graph (ESIG). Event interaction graph is a refinement of event flow graph (EFG) [1]. EIG contains events that interact with the business logic of the GUI application. Event Semantic Interaction (ESI) is used to identify set of events that need to be tested together in multi-way interactions [3] and it is more useful when partitioning the events according to its functionality.

This paper is organized as follow: Section 2 provides some techniques, algorithms used to generate test cases, a method to repair the infeasible test suites are described in section 3, GUI testing on various types of softwares or under different conditions are elaborated in section 4, section 5 describes about testing the GUI application by taking event context into consideration and last section concludes the paper.

## 2. TEST CASE GENERATION

### 2.1. Using GUI Run-Time State as Feedback

Xun Yuan and Atif M Memon [3], used GUI run time state as feedback for test case generation and the feedback is obtained from the execution of a seed test suite on an Application Under Test (AUT).This feedback is used to generate additional test cases and test interactions between GUI events in multiple ways. An Event Interaction Graph (EIG) is generated for the application to be tested and seed test suites are generated for two-way interactions of GUI events. Then the test suites are executed and the GUI's run time state is recorded. This recorded GUI run time state is used to obtain Event Semantic Interaction(ESI) relationship for the application and these ESI are used to obtain the Event Semantic Interaction Graph(ESIG).The test cases are generated and ESIGs is capable of managing test cases for more than two-way interactions and hence forth 2-, 3-,4-,5- way interactions are tested. The newly generated test cases are tested and additional faults are detected. These steps are shown in Figure 1. The fault detection effectiveness is high than the two way interactions and it is because, test cases are generated and executed for combination of events in different execution orders.

There also some disadvantages in this feedback mechanism. This method is designed focusing on GUI applications. It will be different for applications that have intricate underlying business logic and a simple GUI. As multi-way interactions test cases are generated, large number of test cases will be generated. This feedback mechanism is not automated.

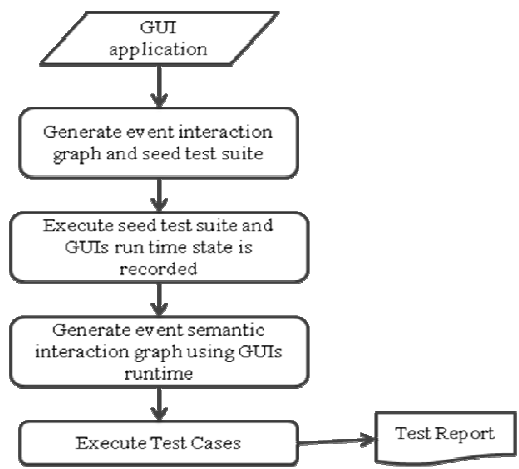

Figure 1. Test Case Generation Using GUI Runtime as Feedback

## 2.2. Using Covering Array Technique

Xun Yuan *et al* [4], proposed a new automated technique for test case generation using covering arrays (CA) for GUI testing. Usually 2-way covering are used for testing. Because as number of events in a sequence increases, the size of test suite grows large, preventing from using sequences longer than 3 or 4. But certain defects are not detected using this coverage strength. Using this technique long test sequences are generated and it is systematically sampled at particular coverage strength. By using covering arrays *t*-way coverage strength is being maintained, but any length test sequences can be generated of at least *t*. A covering array, CA(N; t, k, v), is an N × k array on v symbols with the property that every N × t sub-array contains all ordered subsets of size t of the v symbols at least once.

As shown in Figure 2, Initially EIG model is created which is then partitioned into groups of interacting events and then constraints are identified and used to generate abstract model for testing. Long test cases are generated using covering array sampling. Event sequences are generated and executed. If any event interaction is missed, then regenerate test cases and repeat the steps.

The disadvantages are event partition and identifying constraints are done manually.

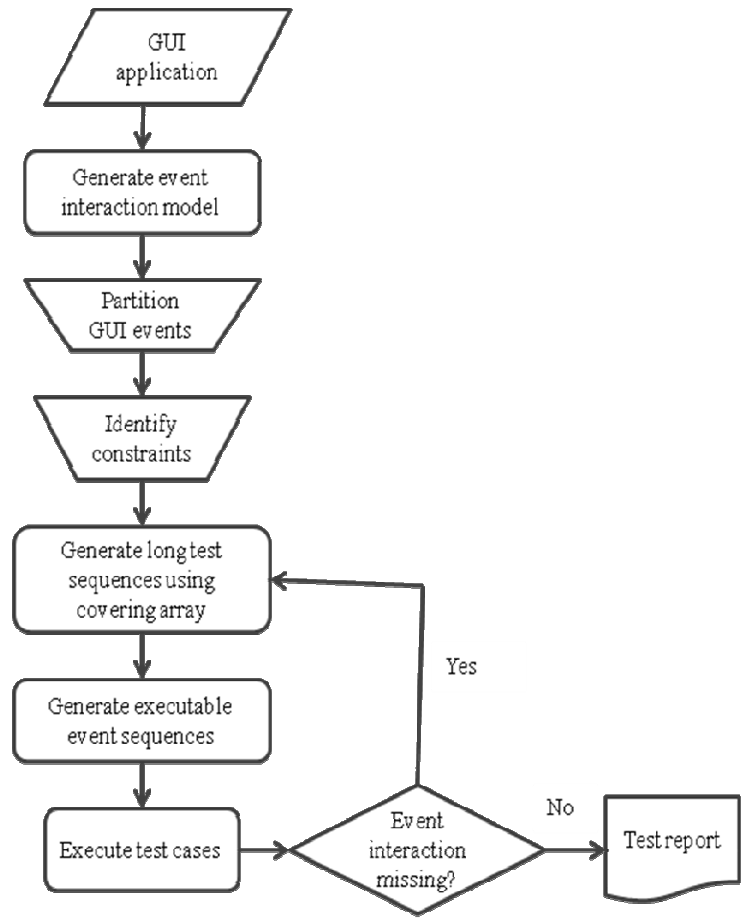

Figure 2. Test Generation Using Covering Array

International Journal of Software Engineering & Applications (IJSEA), Vol.3, No.1, January 2012

## 2.3. Dynamic Adaptive Automated test Generation

Xun Yuan *et al* [5], suggested an algorithm to generate test suites with fewer infeasible test cases and higher event interaction coverage. Due to dynamic state based nature of GUIs, it is necessary and important to generate test cases based on the feedback from the execution of tests. The proposed framework uses techniques from combinatorial interaction testing to generate tests and basis for combinatorial interaction testing is a covering array. Initially smoke tests are generated and this is used as a seed to generate Event Semantic Interaction (ESI) relationships. Event Semantic Interaction Graph is generated from ESI. Iterative refinement is done through genetic algorithm. An initial model of the GUI event interactions and an initial set of test sequences based on the model are generated. Then a batch of test cases are generated and executed. Code coverage is determined and unexecutable test cases are identified. Once the infeasible test cases are identified, it is removed and the model is updated and new batch of test cases are generated and the steps are followed till all the uncovered ESI relationships are covered. These automated test case generation process is shown in Figure 3. This automated test generation also provides validation for GUIs.

The disadvantages are event contexts are not incorporated and need coverage and test adequacy criteria to check how these impacts fault detection.

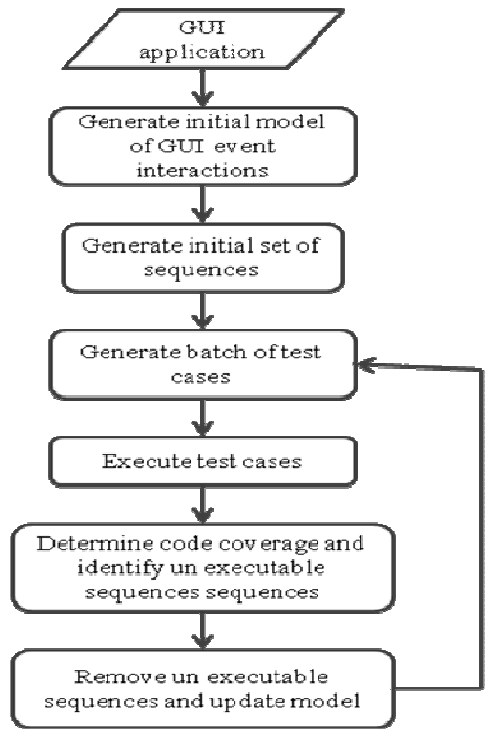

Figure 3. Automated Test Case Generation

## 3. REPAIRING TEST SUITES

Si Huang *et al* [6], proposed a method to repair GUI test suites using Genetic algorithm. New test cases are generated that are feasible and Genetic algorithm is used to develop test cases that provide additional test suite coverage by removing infeasible test cases and inserting new feasible test cases. A framework is used to automatically repair infeasible test cases. A graph model such as EFG, EIG, ESIG and the ripped GUI structure are used as input. The main controller passes

142

these inputs to the test case assembler which then passes the ESIG model to the covering array generator along with the strength of testing. This covering array generator generates an initial set of event sequences. The covering array information is send to test case assembler and it assembles this into concrete test cases. These are passed back to the controller and test suite repair phase begins. Feasible test cases are returned by the framework once the repair phase is complete.

Genetic algorithm is used as a repair algorithm. An initial set of test cases are executed and if there is no infeasible test cases, it exits and is done. If infeasible test cases are present, it then begins the repair phase. A certain number of iterations are set based on an estimate of how large the repaired test suite will be allowed to grow and for each iteration the genetic algorithm is executed. The algorithm adds best test case to the final test suites. Stopping criteria's are used to stop the iterations.

The advantages are it generates smaller test suites with better coverage on the longer test sequences. It provides feasible test cases. But it is not scalable for larger applications as execution time is high. As GUI ripping is used, the programs that contain event dependencies may not be discovered.

## 4. GUI TESTING ON VARIOUS APPLICATIONS

### 4.1. Industrial Graphical User Interface Systems

Penelope Brooks *et al* [7], developed GUI testing methods that are relevant to industry applications that improve the overall quality of GUI testing by characterizing GUI systems using data collected from defects detected to assist testers and researchers in developing more effective test strategies. In this method, defects are classified based on beizer's defect taxonomy. Eight levels of categories are present each describing specific defects such as functional defects, functionality as implemented, structural defects, data defects, implementation defects, integration defects, system defects and test defects. The categories can be modified and added according to the need. If any failures occur, it is analyzed under which defect category it comes and this classification is used to design better test oracle to detect such failures, better test case algorithm may be designed and better fault seeding models may be designed.

Goal Question Metric (GQM) Paradigm is used. It is used to analyze the test cases, defects and source metrics from the tester / researcher point of view in the context of industry-developed GUI software. The limitations are, the GUI systems are characterized based on system events only. User Interactions are not included.

### 4.2. Community-Driven Open Source GUI Applications

Qing Xie and Atif M. Memon [8], presented a new approach for continuous integration testing of web-based community-driven GUI-based Open Source Software(OSS).As in OSS many developers are involved and make changes to the code through WWW, it is prone to more defects and the changes keep on occurring. Therefore three nested techniques or three concentric loops are used to automate model-based testing of evolving GUI-based OSS. Crash testing is the innermost technique operates on each code check-in of the GUI software and it is executed frequently with an automated GUI testing intervention and performs quickly also. It reports the software crashes back to the developer who checked in the code. Smoke testing is the second technique operates on each day's GUI build and performs functional reference testing of the newly integrated version of the GUI, using the previously tested version as a baseline. Comprehensive Testing is the outermost third technique conducts detailed comprehensive GUI integration testing of a major GUI release and it is executed after a major version of GUI is available. Problems are reported to all the developers who are part of the development of the particular version.

These concentric loops provide resource utilization, errors are caught earlier by inner loops. The flaws that persist across multiple versions GUI-based OSS are detected by this approach fully automatically. It provides feedback. The limitation is that the interactions between the three loops are not defined.

### 4.3. Continuously Evolving GUI-Based Software Applications

Qing Xie and Atif M. Memon [9], developed a quality assurance mechanism to manage the quality of continuously evolving software by Presenting a new type of GUI testing, called crash testing to help rapidly test the GUI as it evolves. Two levels of crash testing is being described: immediate feedback-based crash testing in which a developer indicates that a GUI bug was fixed in response to a previously reported crash; only the select crash test cases are re run and the developer is notified of the results in a matter of seconds. If any code changes occur, new crash test cases are generated and executed on the GUI. Test cases are generated that can be generated and executed quickly and cover all GUI functionalities. Once EIG is obtained, a boolean flag is associated with each edge in the graph. During crash testing, once test cases that cover that particular edge are generated, then the flag is set. If any changes occur, boolean flag for each edge is retained. Test cases are executed and crashes during test execution are used to identify serious problems in the software. The crash testing process is shown in Figure 4. The effectiveness of crash test is known by the total number of test cases used to detect maximum faults. Significantly, test suite size has no impact on number of bugs revealed.

This crash testing technique is used to maintain the quality of the GUI application and it also helps in rapidly testing the application. The drawbacks are, this technique is used for only testing GUI application and cannot used in web applications, Fault injection or seeding technique, which is used to evaluate the efficiency of the method used is not applied here.

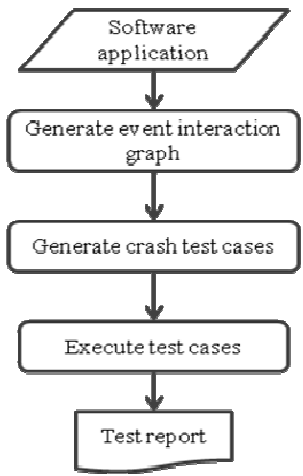

Figure 4. Crash Testing Process

### 4.4. Rapidly Evolving Software

Atif M. Memon *et al* [10], made several contributions in the area of GUI smoke testing in terms of GUI smoke test suites, their size, fault detection ability and test oracle. Daily Automated Regression Tester (DART) framework is used to automate GUI smoke testing. Developers work on the code during day time and DART automatically launches the Application Under Test (AUT) during night time, builds it and runs GUI smoke tests. Coverage and error report are mailed to developer. In DART all the process such as Analyzing the AUT's GUI structure using GUI ripper, Test case generation, Test oracle generation, Test case executor, Examining the

International Journal of Software Engineering & Applications (IJSEA), Vol.3, No.1, January 2012

reports and unsuccessful test cases, Submitting bug reports are automated. GUI smoke test cases and test oracles are generated. Fault seeding is used to evaluate fault detection techniques used. An adequate number of faults of each fault type are seeded fairly.

The disadvantages are Some part of code are missed by smoke tests, Some of the bugs reported by DART are false positive, Overall effectiveness of DART depends on GUI ripper capabilities, Not available for industry based application testing, Faults that are not manifested on the GUI will go undetected

## 5. INCORPORATING EVENT CONTEXT

Xun Yuan *et al* [1], developed a new criterion for GUI testing. They used a combinatorial interaction testing technique. The main motivation of using combinatorial interaction is to incorporate context and it also considers event combinations, sequence length and include all possible event. Graph models are used and covering array is used to generate test cases which are the basis for combinatorial interaction testing.

A tool called GUITAR (GUI Testing Framework) is used for testing and this provides functionalities like generate test cases, execute test cases, verify correctness and obtain coverage reports. Initially using GUI ripper, a GUI application is converted into event graph and then the events are grouped depending on functionality and constraints are identified. Covering array is generated and test sequences are produced. Test cases are generated and executed. Finally coverage is computed and a test adequacy criterion is analyzed.

The advantages are: contexts are incorporated, detects more faults when compared to the previous techniques used. The disadvantages are infeasible test cases make some test cases unexecutable, grouping events and identifying constraints are not automated.

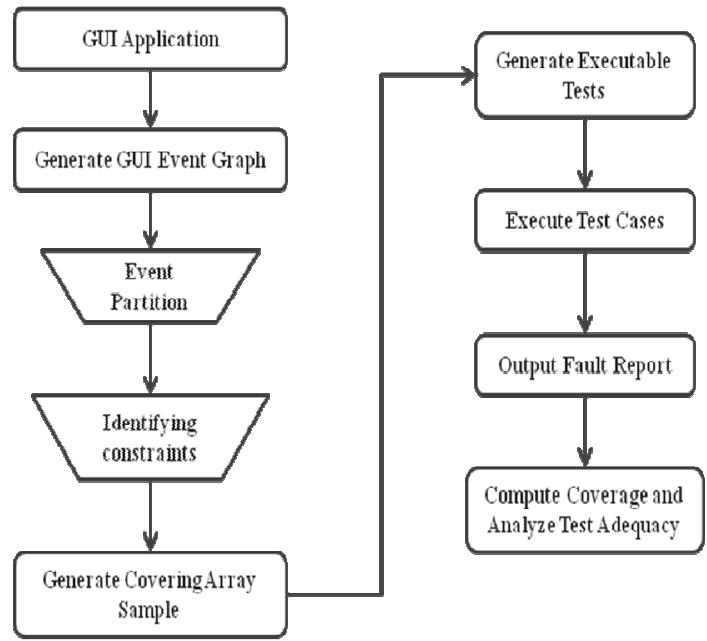

Figure 5. Testing Process

## 6. CONCLUSIONS

In this paper, some of the various test case generation methods and various types of GUI testing adapted for different GUI applications and techniques are studied. Different approaches are being used under various testing environment. This study helps to choose the test case generation technique based on the requirements of the testing and it also helps to choose the type of GUI test to perform based on the application type such as open source software, industrial software and the software in which changes are checked in rapidly and continuously.

## REFERENCES

[1]   Xun Yuan, Myra B. Cohen, Atif M. Memon, (2010) "GUI Interaction Testing: Incorporating Event Context", *IEEE Transactions on Software Engineering*, vol. 99.

[2]   A. M. Memon, M. E. Pollack, and M. L. Soffa, (2001) "Hierarchical GUI test case generation using automated planning", *IEEE Transactions on Software Engineering,* Vol. 27, no. 2, pp. 144-155.

[3]   X. Yuan and A. M. Memon, (2007) "Using GUI run-time state as feedback to generate test cases", in *International Conference on Software Engineering (ICSE),* pp. 396-405.

[4]   X. Yuan, M. Cohen, and A. M. Memon, (2007) "Covering array sampling of input event sequences for automated GUI testing", in *International Conference on Automated Software Engineering (ASE),* pp. 405-408.

[5]   X. Yuan, M. Cohen, and A. M. Memon, (2009) "Towards dynamic adaptive automated test generation for graphical user interfaces", in *First International Workshop on TESTing Techniques & Experimentation Benchmarks for Event-Driven Software (TESTBEDS),* pp. 1-4.

[6]   Si Huang, Myra Cohen, and Atif M. Memon, (2010) "Repairing GUI Test Suites Using a Genetic Algorithm, "in *Proceedings of the 3rd IEEE International   Conference on Software Testing Verification and Validation (ICST)*.

[7]   P. Brooks, B. Robinson, and A. M. Memon, (2009) "An initial characterization of industrial graphical user interface systems", in *ICST 2009: Proceedings of the 2$^{nd}$ IEEE International Conference on Software Testing, Verification and Validation*, Washington, DC, USA: IEEE Computer Society.

[8]   Q. Xie, and A.M. Memon (2006) "Model-based testing of community driven open-source GUI applications", in *International Conference on Software Maintenance (ICSM),* pp. 145-154.

[9]   Q. Xie and A. M. Memon, (2005) "Rapid "crash testing" for continuously evolving GUI- based software applications", in *International Conference on Software Maintenance (ICSM),* pp. 473-482.

[10]  A. M. Memon and Q. Xie, (2005) "Studying the fault-detection effectiveness of GUI test cases for rapidly evolving software", *IEEE Transactions on Software Engineering,* vol. 31, no. 10, pp. 884-896.

[11]  U. Farooq, C. P. Lam, and H. Li, (2008) "Towards automated test sequence generation", in *Australian Software Engineering Conference,* pp. 441-450.

**Authors**

A. Isabella is a Post Graduate research scholar in the department of Software Engineering at Karunya University. Her research areas include GUI Software Testing and Software Engineering.
.

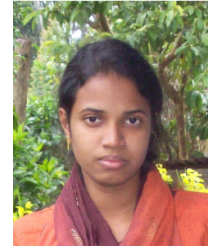

Ms. J. Emi Retna is heading the Computer Technology Centre at KarunyaUniversity. She has eight years of experience. Five years in IT industry and three years in University academics and administation. Her research interests include Java based application development, Enhancing the performance of applications, Software engineering concepts and Quality Metrics.

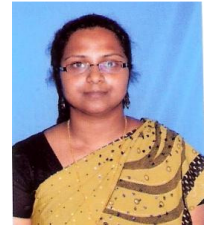